\documentclass[aps,11pt]{revtex4}
\begin{document}
\title{Comment on Phys. Stat. Sol. (b) 236 (2003) 281 paper by A. M. Oles "Orbital ordering and orbital fluctuations in
transition metal oxides"}
\author{R.J. Radwanski}
\homepage{http://www.css-physics.edu.pl}
\email{sfradwan@cyf-kr.edu.pl}
\affiliation{Center for Solid State Physics, S$^{nt}$Filip 5, 31-150 Krakow, Poland,\\
Institute of Physics, Pedagogical University, 30-084 Krakow, Poland}
\author{Z. Ropka}
\affiliation{Center for Solid State Physics, S$^{nt}$Filip 5, 31-150
Krakow, Poland}

\begin{abstract}
We argue that the $^{3}A_{2}$ state considered by Oles in Phys.
Stat. Sol. (b) 236 (2003) 281 for the $d^{2}$ system occurring in
the V$^{3+}$ ion in V$_{2}$O$_{3}$ and LaVO$_{3}$ as well as in
Ti$^{2+}$ ion in TiO and in many other oxides is wrong. The proper
ground state is $^{3}T_{1g}$ - its 9-fold degeneracy is further
split in a crystal by intra-atomic spin-orbit interactions and lattice distortions.
\end{abstract}
\maketitle

Oles in Phys. Stat. Sol. (b) 236 (2003) 281 \cite{bib1} presents in
Fig. 1 excitations spectra for $d^{8}$, $d^{5}$, $d^{2}$ and $d^{3}$
systems. According to Fig. 1b excitations spectra in cubic
transition metal oxides for $d^{2}$ ions have the ground state
$^{3}A_{2}$ and higher states $^{1}T_{2}$, $^{1}E$ and $^{1}A_{1}$.
According to us this ground state is wrong. For the $d^{2}$ system
in the octahedral anion surrounding the ground state is $^{3}T_{1g}$
\cite{bib2,bib3}. The state $^{3}T_{1g}$ is completely different
from the Oles ground state $^{3}A_{2}$ - the latter has 3-fold
degeneracy whereas the former - 9-fold degeneracy. The state
$^{3}A_{2}$ is the orbital singlet whereas $^{3}T_{1g}$ is an
orbital triplet. This difference is of fundamental importance in
modern solid-state physics owing to widely discussed properties of
$V_{2}$O$_{3}$, LaVO$_{3}$ and YVO$_{3}$ not mention TiO or
CrO$_{2}$. Behind these states is completely different physics. By
this Comment we would like to clarify the ground state of the
V$^{3+}$ and Ti$^{2+}$ ions in the octahedral crystal field. Despite
of more than 50 years of intensive studies of, say, V$_{2}$O$_{3}$,
its ground state has not been established yet becoming at present a
subject of the very strong controversy.

The many-electron $^{3}T_{1g}$ state as the ground state of the
$d^{2}$ system occurring in V$_{2}$O$_{3}$ has been calculated by
us for the SCES-02 Conference \cite{bib4}. We have considered the
V$^{3+}$ ion in the octahedral anion surroundings and, unlike
others, with taking into account strong intra-atomic electron
correlations of the intra-atomic nature and the spin-orbit
coupling \cite{bib2,bib3,bib4}. The 9-fold degeneracy of the
$^{3}T_{1g}$ subterm is further split in a crystal by intra-atomic
spin-orbit interactions and lattice distortions \cite{bib2,bib3}.
It is plausible that the $^{3}T_{1g}$ ground octahedral subterm in
a solid compound, where 3$d$ atoms are the full part of the
crystallographic lattice, is in agreement with the ground state of
the 3$d^{2}$ system embedded in the lattice as impurities
\cite{bib5}. Most scientists work in the completely different
description for 3$d$ electrons ignoring the atomic integrity of
the 3$d$ ion, what is visible in no usage of the atomic
many-electron notation. In a recent paper by Horsch et al.
\cite{bib6}, of which Oles is the coauthor, a state $^{3}T_{2}$ is
once mentioned to be the ground state of $V^{3+}$ ions in
$V_{2}O_{3}$. However, there was no explanation for the change of
the ground state compared to the commented paper. The commented
paper was even not mentioned.

\textbf{Note added after the referee reports.} Both referees admit
that "the ground state of the $d^{2}$(t$_{2g}$)-system (Ti$^{2+}$
and V$^{3+}$ ion) is 9-fold degenerate $^{3}$T$_{1}$ level" but
one of them claims that it "is only improper labeling" that
according to the referee "has absolutely no consequence on the
results obtained by Oles". We cannot agree that it is only an
improper labeling. The commented paper was prepared (submission
date: July 1, 2002) at the same time when our submission (May 31,
2002) to the SCES-02 Conference \cite{bib4} has been rejected
(December 23, 2002) by the Publishing Committee as presenting
incorrectly the many-electron $^{3}T_{1g}$ state as the ground
state of $V^{3+}$ ions in $V_{2}O_{3}$. Oles was a member of the
Organizing Committee of the SCES-02 as well as the leading member
of the Editorial Board. Thus, it is not "only improper labeling"
as the referee would like. Oles has worked with the $^{3}A_{2}$
ground state making use of the d$^{8}$ ground state and the
hole-particle symmetry as is written on p. 282, line 14 top of the
commented paper and in his recalled paper \cite{bib7} in the
commented paper as Ref. 8. In fact, the wrong Fig. 1 is directly
taken from that paper.

We also cannot agree with the second point of the referee. A
theory that does not distinguish between the 3-fold degenerate
$^{3}A_{2}$ ground state and the 9-fold degenerate $^{3}T_{1g}$
ground state surely is not physically useful.

At the end we would like to add that we somehow like the Oles
ground state as it is already a many-electron state what we
consider as a large progress towards our understanding within the
quantum atomistic solid-state approach (QUASST) \cite{bib8,bib9}.
Both our and Oles approach contrasts a customary qualitative
consideration with single-electron $t_{2g}$/$e_{g}$ states and/or
with 3$d$ bands of the 1-5 eV width.

\end{document}